\documentclass[prl,aps,twocolumn,longbibliography]{revtex4-2}
\usepackage{amsmath,graphicx,amssymb}
\usepackage{xcolor}
\usepackage[colorlinks=true,citecolor=blue,urlcolor=blue,linkcolor=blue]{hyperref}

\DeclareMathOperator{\erf}{erf}
\DeclareMathOperator{\erfc}{erfc}
\DeclareMathOperator{\erfi}{erfi}
\DeclareMathOperator{\erfcx}{erfcx}

\renewcommand{\vec}[1]{\mathbf{#1}}

\newcommand{\abs}[1]{\lvert {#1}{\rvert}}

\begin{document}

\title{Fractional short-time dynamics in driven quantum gases}
\author{Uri Sharell and Tilman Enss}
\affiliation{Institut für Theoretische Physik, Universität Heidelberg, 69120 Heidelberg, Germany}
\date{\today}

\begin{abstract}
    Quantum gases with short-range attractive interaction tend to form pairs. For time-dependent interaction we find that the pairing amplitude at small separation satisfies a fractional differential equation (FDE).  We derive analytic solutions of the pairing evolution for sudden interaction quenches and power-law drives toward resonant scattering.  We observe universal short-time dynamics governed by a nonrelativistic conformal fixed point at which the momentum distribution exhibits self-similar dynamic scaling, in quantitative agreement with experiment.  At longer times, many-body effects induce relaxation toward an equilibrium state.  In this limit, the FDE turns into a Müller-Israel-Stewart type equation that describes a hydrodynamic attractor approaching equilibrium.
\end{abstract}
\maketitle

\begin{figure}[t]
    \centering
    \includegraphics[width=\linewidth]{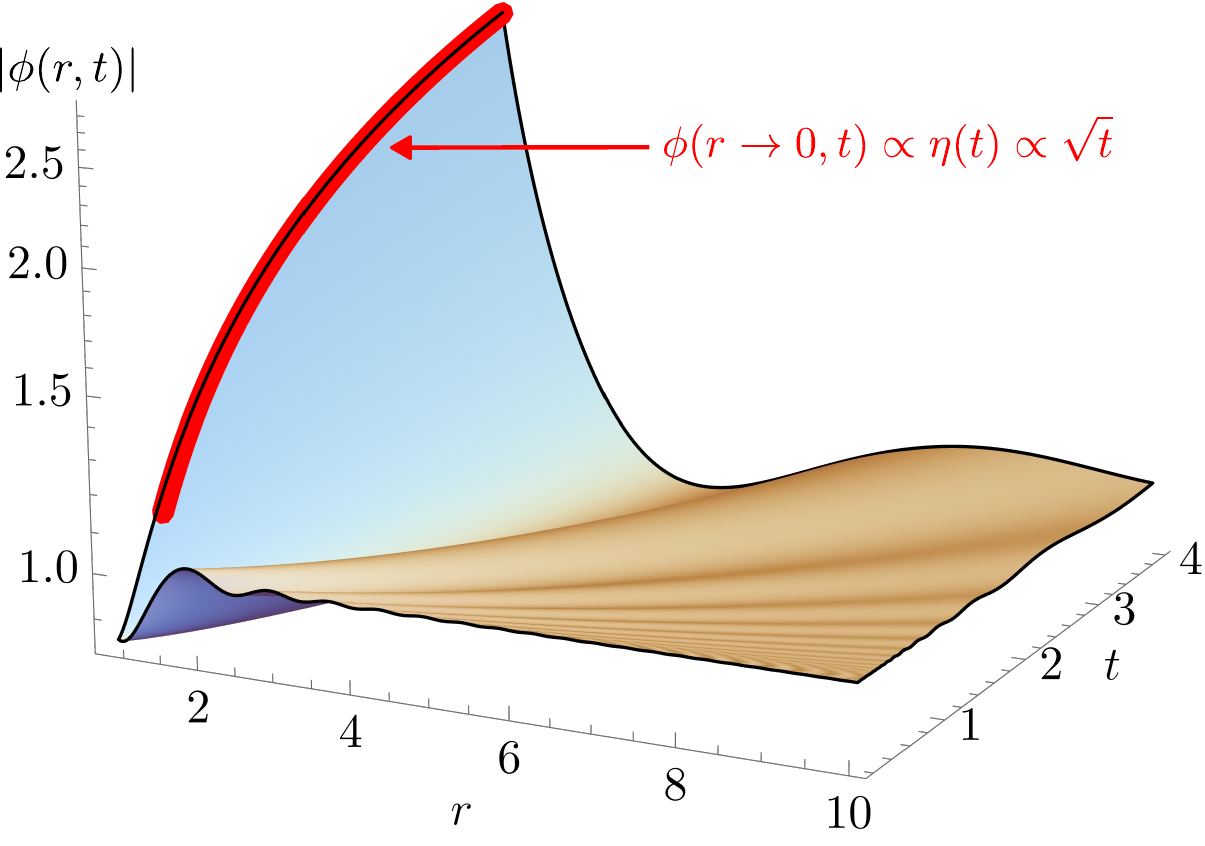}
    \caption{Illustration of the wave function solution for a sudden quench to unitarity as a function of relative distance $r$ and time $t$. At short distance $r \rightarrow 0$, the wavefunction is governed by the pairing amplitude $\eta$ (Eq.~\eqref{eq:bethe}, red boundary line), which exhibits conformal scaling $\eta \propto \sqrt{t}$.}
    \label{fig:psi-vs-r-t-illustration}
\end{figure}

Understanding how isolated many-body quantum systems evolve from far-from-equilibrium initial states toward thermal equilibrium is a central challenge in modern physics \cite{polkovnikov2011, gogolin2016, berges2021}. A key open question is how to connect the early-time dynamics, where rapid entanglement growth quickly renders the quantum state intractable \cite{calabrese2005, kaufman2016}, with a late-time regime where a much simpler hydrodynamic description emerges. Ultracold atomic gases provide an ideal experimental platform to investigate this question, as they offer exceptional control and tunability combined with access to the intrinsic dynamical time and spatial scales \cite{bloch2008, langen2015, ueda2020, makotyn2014, sykes2014, clark2017, liebster2025, huang2019, qi2021maximum, dyke2021, cui2024universal, yi2025quantum, vivanco2025, xie2026, journeaux2026}. In the context of heavy-ion collisions, hydrodynamic models have proven remarkably effective at describing the evolution of the quark-gluon plasma even at surprisingly early times far from equilibrium, a phenomenon attributed to so-called hydrodynamic attractors \cite{heller2015, berges2021, soloviev2022, jankowski2023}. This has recently motivated theoretical predictions of hydrodynamic attractor behavior in ultracold atomic gases \cite{fujii2024, mazeliauskas2026, heller2025early}, which are currently being tested experimentally. At the same time, the early-time behavior generally remains nonuniversal, depending on microscopic details of the system, the preparation of the initial state, and potentially the specific protocol driving the system out of equilibrium.

Here, we argue that in dilute quantum gases, universal short-time behavior emerges when the strength of the interaction changes in time. Since the interaction range is typically much shorter than the particle spacing, the short-time evolution is governed by few-body physics, in particular the pairing dynamics between nearby particles. In this limit, the pair wave function satisfies a free Schrödinger equation with a time-dependent boundary condition as the particle separation approaches $r\to0$. Important thermodynamic observables such as the contact \cite{tan2008energetics, tan2008large} and energy growth \cite{huang2019, qi2021maximum} depend on the wave function at this boundary, hence it is useful to derive an equation of motion specifically for this boundary value, a \emph{boundary Schrödinger equation}. We show that this takes the form of a fractional differential equation (FDE) with a fractional time derivative operator that is nonlocal in time and depends on earlier values of the boundary wave function. As illustrated in Fig.~\ref{fig:psi-vs-r-t-illustration}, the full pair wave function solves the usual Schrödinger equation, a local PDE in space-time, while the boundary value $\eta(t)$ solves the boundary Schrödinger equation, an FDE that is nonlocal in time. This is a complete description and the full, position dependent wave function is easily recovered with the help of the Feynman propagator. 

This approach has several advantages: (i) the FDE for a function of time only is solved more easily than the full Schrödinger equation, and in the following we derive analytic solutions for two dynamical protocols, a sudden change in the interaction and a continuous power-law drive. (ii) For certain thermodynamic observables such as the contact or the evolution of the internal energy, only the boundary wave function is needed. (iii) The space-dependent wave function is easily recovered and yields the evolution of the particle momentum distribution. We show that this short-time dynamics is governed by a conformal fixed point, and the momentum distribution satisfies a self-similar scaling form.

\textit{Fractional differential equation.}
The equation of motion for two quantum particles (bosons or fermions) of mass $m$ in three dimensions in the presence of a time-dependent short-range interaction with scattering length $a(t)$ is given by the Schrödinger equation for the relative wave function \cite{bloch2008},
\begin{align}
    \label{eq:schr}
    i\hbar \partial_t \phi(\vec r,t) 
    = -\frac{\hbar^2}{m}\nabla^2\phi+\frac{4\pi\hbar^2a(t)}{m}\,\delta(\vec r)\partial_r[r\phi].
\end{align}
At short distance $r\to0$ the relative wave function is singular and satisfies the Bethe-Peierls boundary condition,
\begin{align}
    \label{eq:bethe}
    \phi(\vec r,t)
    = \frac{\eta(t)}{\sqrt V}\left[ \frac1r - \frac1{a(t)}\right] + \mathcal O(r).
\end{align}
The prefactor ensures that the wave function is normalized, and we parametrize it in terms of the complex \emph{pairing amplitude} $\eta(t)$.  It determines the singular short-range pair correlations $g_2(r\to0) \propto \abs{\eta(t)}^2/r^2$, and one can read off the contact density \cite{tan2008energetics, tan2008large, qi2021maximum} as
\begin{align}
    \label{eq:contactdef}
    \mathcal C(t) = n^2|4\pi\eta(t)|^2
\end{align}
for fermions or bosons of single-component density $n$ in the noninteracting ground state. As shown in the End Matter, one can derive a Dirichlet-to-Neumann map on the boundary $r \rightarrow 0$ of the Schr\"odinger equation which leads to the fractional differential equation
\begin{align} \label{eq:eomfde}
    D_t^{1/2}\eta(t) = \sqrt{iD_0}\left(1+ \frac{\eta(t)}{a(t)}\right), 
\end{align}
with the diffusion quantum
\begin{align}
    D_0 = \frac\hbar m
\end{align}
and $D^{1/2}_t$ representing a ``half'' derivative with respect to time: Roughly speaking, applying it twice yields the ordinary derivative $D_t^{1/2}\, D_t^{1/2}=d/dt$. More precisely, we introduce the Riemann-Liouville fractional integral, which is defined by generalizing Cauchy's formula for repeated integration to non-integer powers  $(0< \alpha< 1)$:
\begin{align} \label{eq:def-RL-integral}
    I_t^\alpha [f(t)] = \frac1{\Gamma(\alpha)}\int_{t_0}^t f(u)\, (t-u)^{\alpha-1}du.
\end{align}
The Caputo fractional derivative used in Eq. \eqref{eq:eomfde} is then defined as the right-inverse $I^\alpha D^\alpha = id$, such that $D^{\alpha}_t = I^{1-\alpha}_t\frac{d}{dt}$. Explicitly, we have 
\begin{align}\label{eq:def-Caputo-deriv}
    D^{1/2}_t \eta(t) = \frac{1}{\sqrt{\pi}} \int_{t_0 = -\infty}^t \frac{\eta'(u)}{\sqrt{t-u}} du  .
\end{align}
The fractional derivative is a nonlocal operator in time which takes into account the entire history of the function, i.e., $t_0 = - \infty$ in Eq. \eqref{eq:def-Caputo-deriv}. The power $\alpha = 1/2 = 1/z$ in our case is a direct consequence of the nonrelativistic Schr\"odinger symmetry $\omega(k) \propto \abs{k}^z$ \cite{nishida2007nonrel}. An equivalent integral equation of motion was previously derived from the Lippmann-Schwinger equation \cite{qi2021maximum}, and solved to leading order in time. Similarly, \cite{journeaux2026} experimentally analyzed the two-body contact dynamics in a Bose gas and derived a similar integral expression including the effects of a non-uniform zero mode. Our identification and simple derivation of these types of equations as a fractional differential equation allows us to give complete analytic solutions to this problem for two general classes of driving protocols of the scattering length $a(t)$.

\textit{Quench solution.}
Consider a single pair initially in equilibrium at a scattering length $a(t<0)=a_0<0$ such that $\eta(t\leq0)=\eta_0=-a_0$. Since $D^{1/2}_t[\eta_0] =0$, we may as well choose $t_0 = 0$ for the operator $D_t^{1/2}$ rather than $t_0 = - \infty$. At time $t=0$, a quench is performed to a new constant scattering length $a(t>0)=a$.  In this case, the equation of motion \eqref{eq:eomfde} is solved by Laplace transform $\mathcal L[\eta(t)]=\tilde\eta(s)$,
$\mathcal L[D_t^{1/2}\eta(t)]=s^{1/2}\tilde\eta(s) - \eta_0/\sqrt{s}$: we find
\begin{align}
    \tilde\eta(s)& = \frac{\eta_0/\sqrt{iD_0s}+1/s}{s^{1/2}/\sqrt{iD_0}-1/a}\,, \\
    \label{eq:etasol}
    \eta(t) &= -a+(\eta_0+a) \erfcx\left(-\sqrt{iD_0t}/a\right).
\end{align}
The scaled error function $\erfcx(z)=(1-\erf z)\exp(z^2)=E_{1/2}(-z)$ can also be expressed by the one-parameter Mittag-Leffler function, which is an eigenfunction of the fractional derivative and thus generalizes the exponential function in fractional calculus. In particular, a quench from noninteracting ($\eta_0=0$) to resonant scattering $1/a=0$ (unitarity) yields nonrelativistic conformal scaling at all times (cf.~\cite{maki2020far, maki2022dynamics}),
\begin{align}
    \label{eq:conformal}
    \eta(t)=\frac2{\sqrt\pi}\sqrt{iD_0t} \qquad (\text{$1/a\to0$}).
\end{align}

The Schr\"odinger symmetry directly leads to the diffusive growth of the coherence length $\ell^2 \propto t$ in the short-time dynamics, analogous to the speed limit observed in the phase coarsening dynamics of Bose gases \cite{martirosyan2025, liang2026}. We note in passing that this is an example of complex scaling flows, where the dynamic evolution becomes stationary when expressed in new, \emph{complex} space coordinates $\vec r'=\vec r/\sqrt{it}$ \cite{enss2022}.

\begin{figure}
    \includegraphics[width=0.96\linewidth]{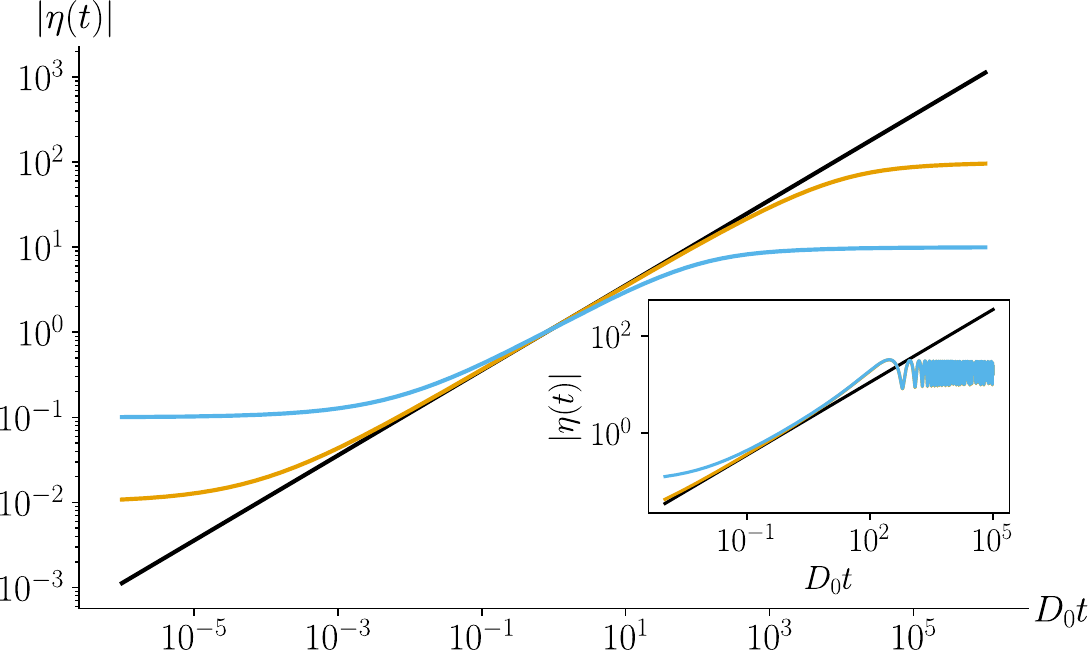}
    \caption{Conformal fixed point scaling in quench dynamics \eqref{eq:etasol}.  Modulus of the pairing amplitude $|\eta(t)|$ vs time after quench from weak ($a_0<0$) to strong ($a<0$) attractive interaction.  (Black) Quench from noninteracting $a_0=0$ to unitary $a=-\infty$ shows conformal scaling $|\eta(t)|=2\sqrt{D_0t/\pi}$.  (Orange) Quench from weak interaction $a_0=-0.01$ to strong ($a=-100$) is first attracted to fixed-point scaling, but eventually deviates toward its equilibrium value $|\eta(t\to\infty)|=-a$.  (Blue) Same for moderate $a_0=-0.1$ to $a=-10$. Inset: Cyclic attractor for quench to repulsive scattering length $a>0$.  (Orange) Quench from weak attraction ($a_0=-0.01$) to moderate repulsion $a=10$.  (Blue) Quench from moderate attraction ($a_0=-0.1$) to moderate repulsion $a=10$.  For both initial conditions the evolution converges toward the cyclic attractor curve \eqref{eq:eta_attr}.}
    \label{fig:eta_attractor}
\end{figure}

\textit{Conformal fixed point scaling.}
The evolution of the pairing amplitude $\eta(t)$ exhibits conformal fixed point scaling.  For a quench from weak to strong attractive interaction, the pairing amplitude is attracted toward the scaling solution \eqref{eq:conformal} and quickly forgets about the initial value $\eta_0$ (see Fig.~\ref{fig:eta_attractor}).  For a quench to resonance, the conformal scaling will continue indefinitely for a single pair, or until the many-body scale is reached.  For a quench to a finite scattering length, the pairing amplitude eventually deviates from the conformal attractor and relaxes toward an equilibrium value $\eta_\text{eq}=-a$ on the interaction time scale $a^2/D_0$.

This quench protocol thus connects an early-time attractor toward conformal fixed point scaling with a late-time crossover toward thermalization.  The approach to equilibrium at $a<0$ is governed by an attractor solution that starts conformally at early times ($\eta_0=0$),
\begin{align}
    \label{eq:eta_attr}
    \eta_\text{attractor}(t)=-a[1-\erfcx(-\sqrt{iD_0t}/a)].
\end{align}
For repulsive scattering $a>0$, there is a bound state that is populated following the quench \cite{corson2015bound, drescher2021}: in Fig.~\ref{fig:eta_attractor} (inset) the attractor solution spontaneously starts oscillating, in distinction to the cyclic attractor induced by a periodic drive \cite{mazeliauskas2026}. Remarkably, in the presence of a bound state, the pairing amplitude $\eta(t)$ can grow faster than conformal scaling $\eta \propto \sqrt{t}$ (in black), which shows that the previously predicted maximum energy growth \cite{qi2021maximum} only holds to leading order in time and can be violated at intermediate times.

\begin{figure}[t]
    \includegraphics[width=1.\linewidth]{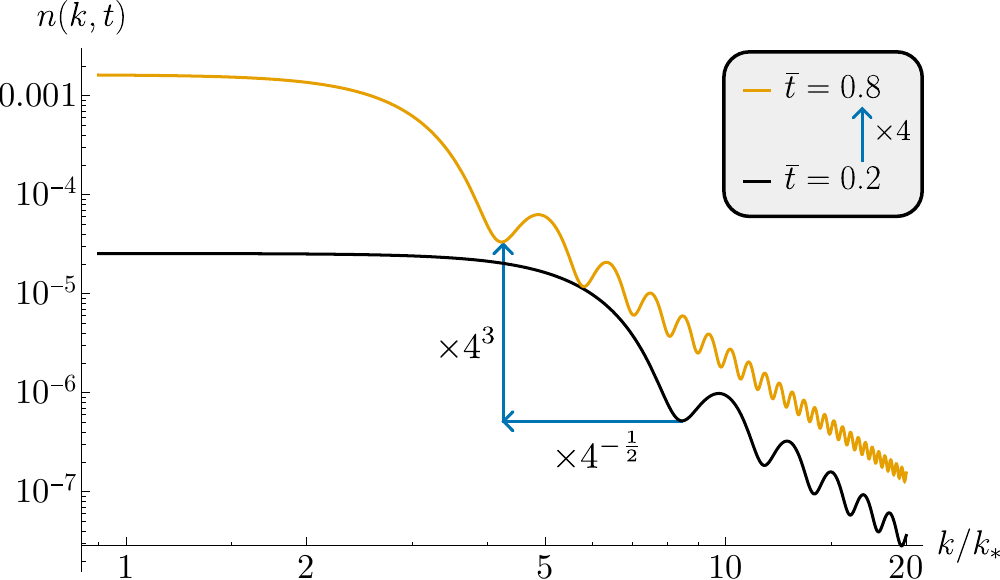}
    \caption{Self-similar momentum distribution $n(k,t)$ vs momentum after a quench to the scattering resonance at times $\bar{t} = D_0k_*^2 t=0.2$ (black) and $\bar{t}=0.8$ (orange), where the reference scale is defined through the characteristic momentum $k_*^3 = 6 \pi^2n$ in analogy with the Fermi momentum.  One observes conformal scaling with a unique scaling function $f_s$ \eqref{eq:nktscaling}, which at the later time (orange) is shifted left by a factor $2$ ($\beta=1/2$) and up by a factor $64$ ($\alpha=3$).  The coefficient of the $k^{-4}$ tail \eqref{eq:nktail} defines the contact density $\mathcal C(t)$.}
    \label{fig:momdist}
\end{figure}

\begin{figure}
    \centering
    \includegraphics[width=.95\linewidth]{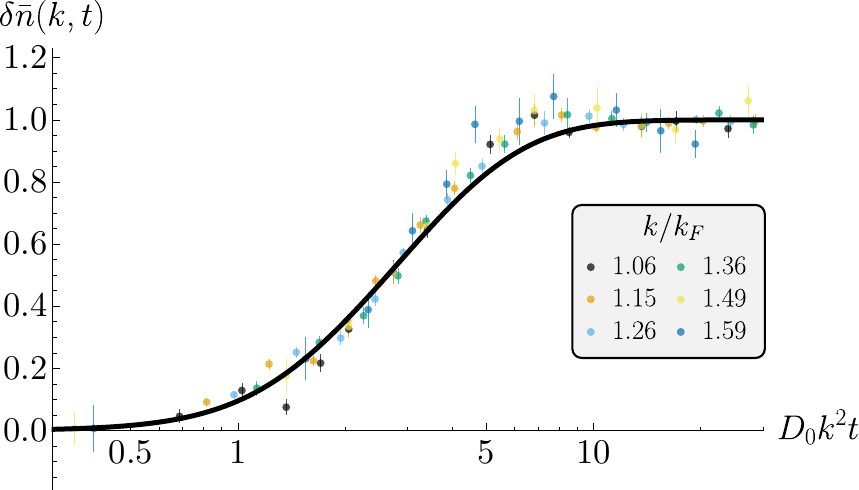}
    \caption{Experimental data showing the normalized change in the momentum distribution due to interaction from \cite{yi2025quantum}. The collapse of different $k$ curves confirms the scaling form \eqref{eq:nktscaling}. The full black line is the theoretical prediction computed at $T/T_F=0.26$, $k/k_F = 1.26$ and $\tau/\tau_F = 0.9$ from Eq.~\eqref{eq:n-exp-def}.}
    \label{fig:exp-comparison}
\end{figure}

\textit{Self-similar scaling of the momentum distribution.}
From the knowledge of $\eta(t)$, the full wave function $\phi(\vec r,t)$ in position space and its Fourier transform $\tilde \phi(\vec k,t) = \int d^3\vec{r}\, e^{i\vec k\cdot \vec r}\,\phi(\vec r,t)$ are reconstructed via the Lippmann-Schwinger equation \cite{qi2021maximum} and we find
\begin{align}\label{eq:phitilde-unevaluated}
    \tilde \phi(k,t)
    = \tilde \phi(k,0) + \frac{4\pi iD_0}{V^{1/2}} \int_0^tdu\, G_0(k,t-u)\,\eta(u)
\end{align}
with Feynman propagator $G_0(k,t)=\exp(-iD_0 k^2t)$ and volume $V$. Consider a quench from noninteracting ($\eta_0=0$) to interacting particles ($a(t>0)=a$). For the attractor solution \eqref{eq:eta_attr} we obtain
\begin{multline}
    \tilde\phi(k,t)
    = \frac{-4\pi a}{V^{1/2}k^2} \Bigl[ 1-\frac{k^2a^2}{1+k^2a^2} \erfcx(-\sqrt{iD_0t}/a) \\
    \label{eq:phitilde}
    - \frac{1-ka\erfi(\sqrt{iD_0k^2t})}{1+k^2a^2} \exp(-iD_0k^2t)\Bigr];
\end{multline}
its Fourier transform agrees with $\phi(r,t)$ computed in \cite{drescher2021}. For a quench to the scattering resonance $1/a\to0$, this simplifies (cf.~\cite{cui2024universal}, and \cite{sykes2014} for trapped gas)
\begin{multline}
    \tilde\phi_\text{reso}(k,t) = \frac{4\pi}{V^{1/2}k^3}\Bigl[\frac2{\sqrt\pi}\sqrt{iD_0k^2t} \\
    - \erfi(\sqrt{iD_0k^2t}) \exp(-iD_0k^2t) \Bigr].
\end{multline}
The single-particle momentum distribution due to excitation of pairs follow as $n(k,t) = n^2V |\tilde\phi(k,t)|^2$. It satisfies a conformal, self-similar scaling form
\begin{align}
    \label{eq:nktscaling}
    & n(k,t) = t^\alpha \tilde f_s(kt^\beta,1/ka)
    \xrightarrow{1/a\to0}
    t^\alpha f_s(kt^\beta)
\end{align}
with $\alpha=3$, $\beta=1/z=1/2$ (see Fig.~\ref{fig:momdist}).  The value of $\beta=1/2$ follows from the Schr\"odinger symmetry of the Feynman propagator in Eq.~\eqref{eq:phitilde} and the scaling form $\eta(t)=\sqrt{D_0t}f_\eta(\sqrt{D_0t}/a)$, and is manifest in the collapse of the experimental data in Fig.~\ref{fig:exp-comparison}. The exponent $\alpha=d = 3$ is most easily seen from $\tilde\phi(k=0,t)\sim (D_0t)^{3/2}$. The large-momentum tail $\tilde\phi(k\to\infty,t)=4\pi\eta(t)/(V^{1/2}k^2)$ implies
\begin{align}
    \label{eq:nktail}
    n(k\to\infty,t) = \frac{n^2|4\pi\eta(t)|^2}{k^4}
    = \frac{\mathcal C(t)}{k^4},
\end{align}
which defines the contact density in Eq.~\eqref{eq:contactdef} above. At the conformal point the contact density $\mathcal{C}(t)\sim t^{\alpha-4\beta}=t$ grows linearly in time \cite{sykes2014, corson2015bound, qi2021maximum, enss2022, cui2024universal}.  While the large-momentum tail encodes the instantaneous pairing amplitude, the zero-momentum limit of Eq.~\eqref{eq:phitilde} yields the time-integrated pairing amplitude that quantifies the growth of coherence.  Note that the values of the exponents reflect neither the conservation of particle number ($\alpha=d\beta$) nor of energy ($\alpha=(d+z)\beta$): out of a reservoir of unperturbed particles, the quench creates a growing number of pair excitations, which saturates only when the many-body scale is reached.

\textit{Relaxation and Müller-Israel-Stewart (MIS) equation.}
In a many-body system of finite density and temperature, the pairing dynamics is more complicated. The pairing growth following a quench is balanced by the decay of pairs in medium on a characteristic time scale $\tau$. By including a phenomenological decay term in the Feynman propagator, $G(k,t) = \exp(-iD_0k^2t) \exp(-t/\tau)$, the evolution equation for $\eta$ is modified as
\begin{align}\label{eq:tempered}
    \mathbb{D}_t^{1/2,1/\tau}\eta(t)
    = \sqrt{iD_0} \left(1+\frac{\eta(t)}{a(t)}\right),
\end{align}
where $\mathbb{D}_t^{\alpha,\gamma}[f]= e^{-\gamma t} D_t^\alpha[e^{\gamma t} f]$ is the \emph{tempered fractional derivative,} which appears in the context of diffusion models \cite{sabzikar2015}. Physically, the exponential terms limit the memory of the nonlocal derivative to the decay time $\tau$. For a system initially interacting $a(t\leq0)=a_0$ with equilibrium value $\eta(t\leq0)=\eta_0=(1/\sqrt{iD_0\tau}-1/a_0)^{-1}$, we can explicitly compute the contribution from the infinite past up to $t=0$ as $\pi^{-1/2}e^{-t/\tau} \int_{-\infty}^0 du \, \partial_u(\eta_0 e^{u/\tau})/\sqrt{t-u} = \eta_0/\sqrt{\tau} \erfc{\sqrt{t/\tau}}$. As expected, the system exponentially forgets about the initial condition $\eta_0$. This ultimately gives
\begin{align}\label{eq:FDE-with-damping}
     \mathbb{D}_{t,t_0=0}^{1/2, 1/\tau} \Bigl[ \eta(t)-\eta_0\erfc\sqrt{t/\tau}\Bigr]
    = \sqrt{iD_0} \Bigl(1+\frac{\eta(t)}{a(t)} \Bigr),
\end{align}
with the fractional derivative now only integrating from $t_0 = 0$. For quenched interaction $a(t>0) = a$, it is possible to rewrite the tempered FDE \eqref{eq:FDE-with-damping} as an ODE (see End Matter for details) to obtain
\begin{align}
    \label{eq:ODEdamped}
    \dot\eta
    & = \Bigl( \frac{iD_0}{a^2} - \frac1\tau \Bigr) \eta 
    + \Bigl( \frac{iD_0}a+\sqrt{\frac{iD_0}\tau} \Bigr) \\
    & + \sqrt{iD_0}\Bigl(1+\eta_0\bigl(\frac1a-\frac1{\sqrt{iD_0\tau}}\bigr)\Bigr) \Bigl(\frac{e^{-t/\tau}}{\sqrt{\pi t}}-\frac{\erfc\sqrt{t/\tau}}{\sqrt\tau} \Bigr).\notag 
\end{align}
The first term linear in $\eta$ leads to oscillation with bound-state energy and damping, the second, constant term sets the quench parameters, while the third term is explicitly time dependent and makes the ODE non-autonomous. At short times $ t/\tau\ll1$, the third term gives rise to the conformal fixed point scaling $\eta(t)\sim\sqrt t$, whereas its contribution is suppressed for $ t/\tau\gtrsim1$. This accounts for the exponential loss of information about the initial condition $\eta_0$ and turns the FDE into an autonomous ODE of MIS type that governs hydrodynamic attractor behavior \cite{heller2015, berges2021, soloviev2022, jankowski2023, fujii2024, mazeliauskas2026}. We find the explicit attractor solution
\begin{align}
    \label{eq:etasoldamp}
    \eta(t)
    & = \frac{1-\chi(t)}{1/\sqrt{iD_0\tau}-1/a} + \chi(t) \eta_0 
    \qquad \text{with}\\
    \chi(t) & =\frac{a\erfc\sqrt{t/\tau} + \sqrt{iD_0\tau}\; e^{-t/\tau} \erfcx\Bigl(-\sqrt{iD_0t}/a\Bigr)}{a+\sqrt{iD_0\tau}},
    \notag
\end{align}
which covers the full evolution from the initial conformal attractor toward the final equilibrium value as $t\gg\tau$.

A recent experiment \cite{yi2025quantum} measured the evolution of the momentum distribution following a quench from an ideal Fermi gas toward resonance. For this case we find the solution $\eta(t)=\sqrt{iD_0\tau}\,\erf\sqrt{t/\tau}$, which interpolates between $\sqrt t$ growth at short times and exponential saturation to the equilibrium value determined by the decay time $\tau$. The change in the momentum distribution due to interaction follows from the pair wave function
\begin{multline}
    \label{eq:pairwavefct}
    \tilde \phi(k,t)
    = \frac{4\pi(iD_0\tau)^{3/2}}{V^{1/2}(1+iD_0k^2\tau)}
    \Bigl[ \erf\sqrt{t/\tau} \\ - \exp(-iD_0k^2t-t/\tau) \frac{\erf\sqrt{-iD_0k^2t}}{\sqrt{-iD_0k^2\tau}} \Bigr].
\end{multline}
For early times in the conformal scaling regime this rises as $t^{3/2}$ and saturates for times $t\gtrsim\tau$ toward equilibrium with contact density $\mathcal C=n_\uparrow n_\downarrow |4\pi\eta|^2=0.09(\tau/\tau_F)k_F^4$. By matching this to the computed and measured value $\mathcal C=0.085(5)k_F^4$ at $T/T_F=0.26$ \cite{enss2011, rossi2018, mukherjee2019, yi2025quantum} we obtain $\tau=0.95(5)\tau_F$; this allows us to predict $n(k,t)$ within our model without free parameters. To identify signatures of our two-body quench solution in the many-body experimental data \cite{yi2025quantum}, we must take into account that initial momenta before the quench are populated according to a Fermi distribution. As explained in the End Matter, at low temperature this effectively averages $n(k,t)$ over the scale of the Fermi momentum $k_F$. As shown in Fig.~\ref{fig:exp-comparison}, even this simple model shows remarkable quantitative agreement with the experimentally measured growth of the momentum distribution.

\begin{figure}
    \centering
    \includegraphics[width=\linewidth]{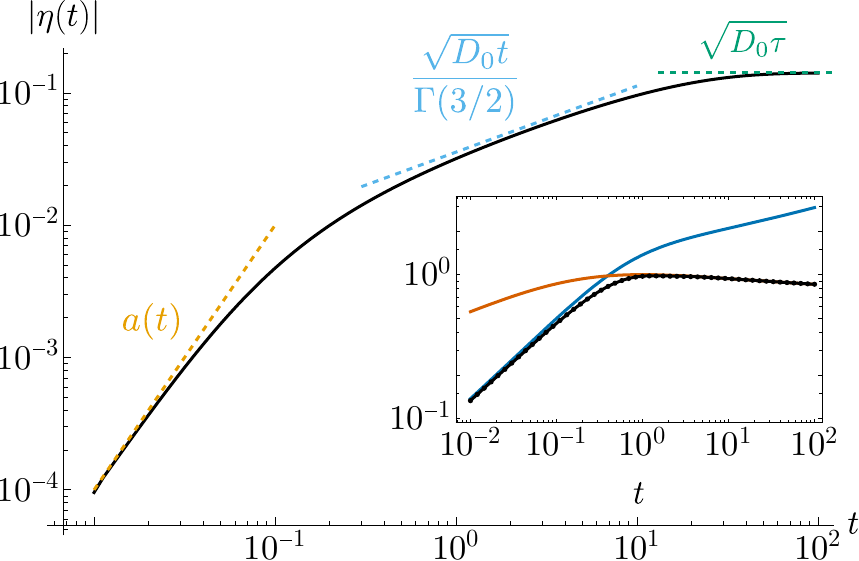}
    \caption{Dynamics of the pairing amplitude $\eta$ for a power-law drive of scattering length $a(t) = - t^2$, with $D_0 = 10^{-3}, \tau = 20$. The black curve shows the solution of the tempered FDE \eqref{eq:FDE-with-damping} obtained with a numerical FDE solver \cite{garrappaCodefde, garrappaNumericalSolutionFractional2018, diethelm1998fracpece}. The different asymptotic scaling forms $\eta  \rightarrow  \min( a(t), \frac{2}{\sqrt{\pi}}\sqrt{D_0t}, \sqrt{D_0\tau})$ are shown by the yellow, blue, and green dashed lines, respectively. Inset: Solution for a different choice of parameters ($a(t) = t^{0.2}, \tau = 0.5, D_0 = 1$), where the black dots show the numerical solution of the tempered FDE and the black line the analytic solution evaluated as a truncated power series from \eqref{eq:powerlaw-sol-analytic-app}. The blue and red lines show the derived simple approximate forms for early \eqref{eq:powerlaw-early-approx-sol} and late times \eqref{eq:powerlaw-late-approx-sol}.}
    \label{fig:powerlaw}
\end{figure}

\textit{Power-law drive.}
Finally, we consider a more general approach to unitarity with a power-law drive $a^{-1}(t) =  A t^{-B}$ with constant prefactor $A$ as in \cite{qi2021maximum}, while setting $\eta_0 = 0$ for simplicity. As derived in the End Matter, we can write the full analytic solution of \eqref{eq:eomfde} for $\lambda = 1/2-B >0$ as
\begin{align}\label{eq:power-law-sol}
    \eta(t) &= \frac{2}{\sqrt{\pi}}\sqrt{i D_0 t}\, E_{1/2,2 \lambda,2 \lambda}\left(\frac{\sqrt{i D_0 t}}{ a(t)}\right),
\end{align}
where $E_{\alpha,\beta,\gamma}(z)$ denotes the Kilbas-Saigo function of Mittag-Leffler type first defined as a power series in \cite{kilbasiMittaglefflerTypeFunction1996} as solutions to certain Abel-Volterra equations \cite{kilbasTheoryApplicationsFractional2006}. The case $\lambda<0$ can also be solved and is discussed in the End Matter. Inspired by a conjecture in \cite{lofti2021}, we find a very useful approximation (for any $\lambda$) to the full solution,
\begin{align} \label{eq:powerlaw-early-approx-sol}
    \eta(t) \approx 
     \left(\frac{\sqrt{\pi}/2}{\sqrt{i D_0 t}}- \frac{1}{a(t)} \right)^{-1}.
\end{align}
This approximation is equivalent to substituting $2/\sqrt{\pi} D_t^{1/2}\eta\approx \eta /\sqrt{t} $ in the original equation, which is exact for the special case $a(t) \propto \sqrt{t}$, as well as for the quench to unitarity $a \rightarrow \infty$, recovering the scaling solution \eqref{eq:conformal}. This form also matches the expectation from simple scaling arguments, assuming a solution of the form $\eta \propto t^\kappa$ such that $D_t^{1/2}\eta  \propto t^{\kappa - 1/2} \propto  \eta/\sqrt{t} \propto (1 + \eta/a)$. The asymptotics at early and late times can therefore be deduced and summarized as $\abs{\eta} \rightarrow \min(2\sqrt{iD_0 t/\pi}, a(t))$: when the scattering length is larger than the $\sqrt{t}$ scaling, the pairing amplitude simply follows the quench solution $\eta_\text{conf} = 2\sqrt{D_0t/\pi}$, whereas for an adiabatic drive of the scattering length the system remains in equilibrium and the pairing amplitude follows its instantaneous equilibrium value $\eta(t) = -a(t)$. Equation \eqref{eq:powerlaw-early-approx-sol} also shows how for repulsive scattering $a> 0$, at intermediate times the pairing amplitude may violate the maximum growth predicted in \cite{qi2021maximum}, $\eta > \eta_\text{conf}$, as previously seen in the discussion of Fig.~\ref{fig:eta_attractor}. The tempered FDE can also be analytically solved by a power series of generalized hypergeometric functions (see End Matter). For early times $t \ll \tau$, the tempered FDE is equivalent to the untempered case \eqref{eq:power-law-sol}, while for $t \gg \tau$ an approximate form of the solution can be derived as
\begin{align} \label{eq:powerlaw-late-approx-sol}
    \eta(t) \approx \left(\frac{1}{\sqrt{i D_0 \tau}}- \frac{1}{a(t)} \right)^{-1}.
\end{align}
This form is equivalent to approximating $\mathbb{D}^{1/2, 1/\tau}\eta = e^{-t/\tau} D^{1/2}(e^{t/\tau} \eta) \approx \eta/\sqrt{\tau}$, which becomes exact in the limit $t/ \tau \rightarrow \infty$. In summary, the evolution of the pairing amplitude with a power-law drive of the scattering length $a(t) \propto t^B$ will approximately follow $\abs{\eta} \rightarrow \min(\frac{2}{\sqrt{\pi}} \sqrt{D_0 t}, a(t), \sqrt{D_0 \tau})$, as further illustrated in Fig.~\ref{fig:powerlaw}. The pairing amplitude closely follows and interpolates between the two approximate solutions \eqref{eq:powerlaw-early-approx-sol} and \eqref{eq:powerlaw-late-approx-sol} as shown in the inset of Fig.~\ref{fig:powerlaw}. These approximations further provide a simple starting point to qualitatively analyze systems with arbitrary scattering length drives.

\textit{Conclusion.}
We have shown that the short-time dynamics of a dilute quantum gas after a rapid increase in interaction satisfies a fractional differential equation and is universal: for different initial conditions the pairing amplitude is attracted toward interaction-driven, conformal fixed point scaling. At later times in the many-body regime, local dissipation leads to thermalization and information loss as modeled by a tempered FDE. The predicted self-similar scaling of the momentum distribution after a quench is confirmed by a recent experiment over the full range from early to late times. In this limit, the tempered FDE takes the form of a Müller-Israel-Stewart equation describing a hydrodynamic attractor for $\eta$.
In future work it will be interesting to determine the relaxation time $\tau$ from a microscopic calculation of the decay of pairs in medium \cite{nishida2019, enss2019bulk, hofmann2020, johansen2024, enss2024, dizer2024}. Our approximate solution \eqref{eq:powerlaw-early-approx-sol} provides starting points for arbitrary drives and even complex $a(t)$ modeling driven-dissipative systems \cite{hudsonsmith2015}.

The authors thank Shujin Deng for sharing their experimental data and J\"urgen Berges, Gabriel Denicol, Keisuke Fujii, Thomas Gasenzer, Giuliano Giacalone, Gia\-como Gori, Micha\l\ Heller, Aleksas Mazeliauskas, Micha\l\ Spali\'nski, Raju Venugopolan, and Clemens Werthmann for discussions.  This work was supported by the DFG (German Research Foundation) under Project No.~273811115 (SFB 1225 ISOQUANT) and under Germany’s Excellence Strategy EXC2181/1-390900948 (the Heidelberg STRUCTURES Excellence Cluster).

\appendix
\section{END MATTER}
\subsection{Derivation of FDE}
Using the Caputo definition for a fractional derivative ($0<\alpha<1)$ $D_{t,t_0}^{\alpha} f = \Gamma(1-\alpha)^{-1} \int_{t_0}^t du  f'(u)/(t-u)^{\alpha}$, it is easy to show for $t_0 = - \infty$ that the properties of Fourier transforms on derivatives generalize to $\mathcal{F}[D_{t}^{\alpha}[f]] = (i \omega)^\alpha \mathcal{F}[f]$ \cite{podlubnyfractional1999}. Consider a smooth and bounded function $u(t,r)$ that satisfies 
\begin{align} \label{eq:harmonic-extension}
    \Bigr(\frac{i}{D_0}\partial_t + \partial_r^2 \Bigr)u = 0,  \quad 
    u(t,r=0) = f(t)
\end{align}
with $D_0 = \hbar/m$. Enforcing the boundary conditions at $r=0$ and $r \rightarrow \infty$, the spatially bounded solution is given in the frequency domain as $\hat{u}(\omega,r) = \hat{f}(\omega) e^{-r \sqrt{\omega/D_0}}$. Taking the derivative with respect to $r$ then gives us the relation
\begin{align}\label{eq:caffarelli-extension-app}
    -\partial_r \hat{u}\rvert_{r=0}= \sqrt{\frac{\omega}{D_0}} \hat{u}\rvert_{r=0} =  \frac{1}{\sqrt{i D_0}} \mathcal{F}[D^{1/2}_{t} u\rvert_{r=0}].
\end{align}

This map between Dirichlet and Neumann boundary conditions is a well-known mathematical result that was popularized by Caffarelli and Silvestre to study fractional Laplacian operators \cite{caffarelli2007extension}. The reduced radial wavefunction $u(r,t) = r \phi(r,t)$ for the 3D quantum gas satisfies precisely Eq.~\eqref{eq:harmonic-extension} with the Bethe-Peierls boundary condition $u\rvert_{r \rightarrow 0} = \eta(t)(1- r/a)/\sqrt{V}$. For an initial zero mode $\phi_0=1/\sqrt V$, however, $u = r\phi_0 \sim r/\sqrt{V}$ is unbounded. We therefore introduce an auxiliary, bounded wavefunction as $u_b = u - r/\sqrt{V}$, which still satisfies Eq.~\eqref{eq:harmonic-extension}. Evaluating the boundary conditions $- \partial_ru_b\rvert_{r=0} = (1+ \eta/a)/\sqrt{V}$ and $u_b\rvert_{r=0} = \eta/\sqrt{V}$ and applying the relation \eqref{eq:caffarelli-extension-app} (after an inverse Fourier transform) immediately yields Eq.~\eqref{eq:eomfde} in the main text. The derivation of the tempered FDE \eqref{eq:tempered} is analogous but with an additional damping term $\phi/\tau$ in the Schrödinger equation, which can be absorbed by rescaling $\phi\rightarrow e^{t/\tau}\phi$. Finally, it is easy to show that the integral operator derived in \cite{qi2021maximum}, $\hat{L}\eta \propto \lim_{\varepsilon\to0^+} \left[ 2\eta(t)/{\sqrt\varepsilon} - \int_0^{t-\varepsilon} du \, \eta(u)/(t-u)^{3/2}\,{} \right]$, corresponds up to a constant prefactor to the Riemann-Liouville derivative $^\text{RL}D_t^{1/2} = \frac{d}{dt} I_t^{1/2} \propto \hat{L}$, such that our equations are equivalent.

\subsection{Derivation of ODE}
For a quench to constant scattering length, it is possible to rewrite the tempered FDE \eqref{eq:FDE-with-damping} as an ODE, by applying the fractional derivative on both sides and substituting the original equation. First, note that Riemann-Liouville (RL) and Caputo derivatives are interchangeable for functions with zero initial condition $f(0) = 0 = \eta(0)  -\eta_0 \erfc{(0)}$, such as in Eq. \eqref{eq:FDE-with-damping}, so that we can use the composition property of the RL derivative $D^{1/2}[D^{1/2}[\eta]] = D^{1}[\eta] = \dot{\eta}$ when $I^{1/2}[\eta](0) = 0$ \cite{kilbasTheoryApplicationsFractional2006}, which  is satisfied for non-singular and thus for any physically relevant cases in our system. For the tempered derivative $\mathbb{D}^{1/2,1/\tau} = e^{-t/\tau}D^{1/2} e^{t/\tau}$, the composition rule $\mathbb{D}^{1/2,1/\tau} (\mathbb{D}^{1/2,1/\tau} \eta) =\eta/\tau + \dot{\eta}$ is then easily verified. We rewrite \eqref{eq:FDE-with-damping} as
\begin{align}\label{eq:FDE-deriv-1}
\mathbb{D}^{\frac{1}{2},\frac{1}{\tau}}\eta = \sqrt{i D_0} \left(1+ \frac{\eta}{a}\right) + \mathbb{D}^{\frac{1}{2},\frac{1}{\tau}}\eta_0 \erfc(\sqrt{t/\tau})
\end{align}
and apply the tempered derivative $\mathbb{D}^{1/2,1/\tau}$ on both sides to obtain
\begin{multline} \label{eq:FDE-deriv-2}
    \dot{\eta} + \frac{\eta}{\tau} - \eta_0 \frac{d}{dt}\erfc\sqrt{t/\tau} - \frac{\eta_0}{\tau} \erfc\sqrt{t/\tau}
    \\ = \sqrt{iD_0}\left( \mathbb{D}^{1/2,1/\tau}[1] + \frac{\mathbb{D}^{1/2, 1/\tau}\eta}{a} \right).
\end{multline}
We explicitly evaluate the terms 
\begin{align}
     \frac{d}{dt}\erfc\sqrt{t/\tau} &= - \frac{e^{-t/\tau}}{\sqrt{\pi\tau t}},\label{eq:FDE-deriv-3}\\
    \mathbb{D}^{1/2,1/\tau} \erfc{\sqrt{t/\tau}} &= \frac{e^{-t/\tau}}{\sqrt{\pi t}} -  \frac{\erfc\sqrt{t/\tau}}{\sqrt{\tau}}, \label{eq:FDE-deriv-4} \\
     \mathbb{D}^{1/2,1/\tau}[1] &= \frac{e^{-t/\tau}}{\sqrt{\pi t}} + \frac{1-\erfc\sqrt{t/\tau}}{\sqrt{\tau}}. \label{eq:FDE-deriv-5}
\end{align}
Substituting $\mathbb{D}^{1/2,1/\tau}\eta$ from \eqref{eq:FDE-deriv-1} into \eqref{eq:FDE-deriv-2} along with all the evaluated derivatives \eqref{eq:FDE-deriv-3}-\eqref{eq:FDE-deriv-5} finally yields Eq.~\eqref{eq:ODEdamped} in the main text.

\subsection{Experimental comparison}
To compare our quench solution to experimental data, we must take into account the presence of a Fermi sea at finite density, where scattering particles have nonzero initial momenta. The measured population growth $\delta n$ of single particles at momentum $\vec k$ arises from all possible scattering events from initially occupied states $\vec p$ in the Fermi sea into final states $\vec{k} = \vec p + \vec q$ with momentum transfer $\vec q$ due to pair formation:
\begin{align}
    \label{eq:n-exp-def}
    \delta n_\text{exp}(k,t) = \frac{1}{\bar{n}}\int_{\vec p}f(p) \;n^2V\bigl|\tilde{\phi}(\vec q=\vec k-\vec p,t)\bigr|^2
\end{align}
using the shorthand notation $\int_\vec{p} = \int \frac{d^3 \vec{p}}{(2 \pi)^3}$. Here $\tilde \phi(q,t)$ denotes the pair wave function \eqref{eq:pairwavefct}, $f(p) = (z^{-1} \exp(\beta p^2/(2m)) + 1)^{-1}$ is the Fermi-Dirac distribution with fugacity $z$, and $\bar{n} = \int_{\vec{p}} f(p)$. For the experimental temperature $T/T_F = 0.26$ we numerically evaluate \eqref{eq:n-exp-def} for different times $t$ at experimentally given $k$, which yields the theory prediction (solid curve) in Fig.~\ref{fig:exp-comparison}.

\subsection{Analytic solution for power-law drive}
We consider the fractional integral equation $\eta = gI_t^{1/2}[1+\eta/a(t)]$ for a scattering length $a(t)^{-1} = A t^{-B}$, where we have chosen $\eta_0 = 0 $ for simplicity and introduced $g = \sqrt{iD_0}$ for visual ease. Through repeated substitution we can rewrite this equation as a Neumann series
\begin{align}
    \eta(t) &= g \sum_{n=0}^\infty \hat{K}^n[f], \quad \hat{K}[f]= gA I_t^{1/2}[t^{-B }f], 
\end{align}
where $\hat{K}^0 = id$ and $f = I_t^{1/2}[1] = \sqrt{t}/\Gamma(3/2)$.
The first term in this series is proportional to $t^{1/2}$, while every successive application of the operator $\hat{K}$ increases the power of $t$ by $\lambda = 1/2-B$ and adds a factor of $g A$. From this dimensional analysis we therefore immediately obtain the Ansatz
\begin{align}
    \eta(t) &=g \sum_{n=0}^\infty \eta_n  =g\sqrt t \sum_{n=0}^\infty c_n \left(g A t^\lambda \right)^n. \label{eq:power-law-solution-app}
\end{align}
To determine the coefficients $c_n$, note that each term in the series is related to the next by application of $\hat{K}$ such that $ \eta_{n+1} = \hat{K} \eta_n$. Using the fractional integral of a power law $I_t^{1/2}[t^\nu] = t^{\nu+1/2} \Gamma(1+\nu)/\Gamma(3/2+\nu)$ (for positive or sufficiently irrational $\nu$), we have
\begin{align}
    \eta_{n+1} = c_{n+1} t^{1/2} (g A t^{\lambda})^{n+1} \stackrel{!}{=} c_n I_t^{1/2}[(g A t^{\lambda})^{n+1}] \\
    \iff
    c_{n+1} = c_n \frac{\Gamma\left((n+1)\lambda + 1\right)}{\Gamma\left((n+1)\lambda +\frac{3}{2}\right)}.
\end{align}
The full solution is therefore given by
\begin{align}\label{eq:powerlaw-sol-analytic-app}
    \eta(t) &= g \sqrt{ t} \sum_{n=0}^\infty c_n(g A t^{\lambda})^n =  \frac{\sqrt{i D_0 t}}{\Gamma(3/2)} E_{1/2,2 \lambda,2 \lambda}\left(\frac{\sqrt{i D_0 t}}{ a(t)}\right),\\
    c_n &= \prod_{k=0}^{n} \frac{\Gamma\left(k\lambda + 1\right)}{\Gamma\left(k\lambda +\frac{3}{2}\right)},\quad \lambda = \frac{1}{2}-B, \quad g = \sqrt{iD_0},
\end{align}
where we have identified the function $E_{\alpha,\beta,\gamma}(z)$ as the generalized Mittag-Leffler function with three parameters first introduced by Kilbas and Saigo \cite{kilbasSolutionIntegralEquation1995,kilbasiMittaglefflerTypeFunction1996}. The Kilbas-Saigo function is an entire function \cite{gorenfloGeneralizedMittaglefflerType1998} (i.e., the above series converges) for $\lambda > 0$, or $B < 1/2$. The exact scaling solution \eqref{eq:conformal}, $\eta \propto \sqrt{t}$, as well as the asymptotic form for $a(t) \gg \sqrt{t}$, is immediately recovered from $E_{\alpha, \beta, \gamma}(0) = 1$. Furthermore, we have the property \cite{kilbasiMittaglefflerTypeFunction1996} $E_{\alpha, 1, \gamma} = \Gamma(\alpha \gamma +1 ) E_{\alpha, \alpha \gamma +1}$, where $E_{\alpha, \beta}$ is the standard two-parameter Mittag-Leffler function. This allows us to further recover the finite quench solution for $\eta_0 = 0$ ($B = 0, \lambda = 1/2, A =a^{-1}$) $\eta(t)= \sqrt{i D_0t}E_{1/2,3/2}(\sqrt{i D_0 t}/a) = a  (- 1 + \erfcx(-\sqrt{i D_ 0t}/a))$. The special solution for $a(t) \propto \sqrt{t}$ is also immediately apparent, as then $\lambda = 0$ and $c_n = \Gamma(3/2)^{-(n+1)}$, such that one can resum \eqref{eq:power-law-solution-app} as a geometric series and recover \eqref{eq:powerlaw-early-approx-sol} exactly.

The tempered FDE \eqref{eq:FDE-with-damping} can be solved analogously by a double power series
\begin{align}\label{eq:damped-sol}
    \eta(t) =& g \sqrt{t}e^{-t/\tau}\sum_{n=0}^{\infty}\sum_{k=0}^\infty \frac{(t/\tau)^k}{k!} c_{n,k} (g A t^\lambda)^n,\\
    c_{n,k} =& \prod_{l=0}^{n} \frac{\Gamma(1+l\lambda +k)}{\Gamma(\frac{3}{2}+ l\lambda + k)}.
\end{align}
Note that $c_{n,0} = c_n$ from before. 
It is possible to perform the $k$ sum and generalize \eqref{eq:power-law-solution-app} with a series of generalized hypergeometric functions $_pF_q$,
\begin{align}
    \eta(t) = g \sqrt{t}~e^{-\frac{t}{\tau}} \sum_{n=0}^\infty c_n (gA t^\lambda)^n ~_{n+1}F_{n+1}\big(a_i,b_i; \frac{t}{\tau}\big),
\end{align}
where $i$ is an integer index running $i = 1, \cdots, n+1$, and the coefficients $a,b$ are given by  $a_i = 1 + (i-1) \lambda$ and $b_i = a_i + 1/2$. Since $_p F_q(z\rightarrow 0) = 1$, we recover the untempered FDE solution \eqref{eq:powerlaw-sol-analytic-app} for $t \ll \tau$. For $t \gg \tau$, we can use the fact that the $k$ sum in Eq.~\eqref{eq:damped-sol} can be identified as an expectation value of a random variable with a Poisson distribution $e^{-\mu} (\mu)^k/k!$ and mean $\mu = t/\tau$
\begin{align}
    \eta(t) = g \sqrt{t} \sum_{n=0}^\infty \mathbb{E}[c_{n,K}] (g A t^\lambda)^n.
\end{align}
For large values $\mu = t/\tau \rightarrow \infty$, the Poisson distribution becomes sharply peaked around the mean, so that we may approximate $\mathbb{E}[c_{n,K}] \approx c_{n,t/\tau}$. Using $\frac{\Gamma(\mu+a)}{\Gamma(\mu+b)} \rightarrow \mu^{a-b}$ as $\mu \rightarrow \infty$, we have $c_{n,t/\tau} \rightarrow  \prod_{l=0}^{n} (t/\tau)^{-1/2} = (t/\tau)^{-(n+1)/2}$ and thus
\begin{multline}
    \eta(t)  \approx g \sqrt{t} \sum_{n=0}^\infty 
    \Bigl(\frac{t}{\tau}\Bigr)^{-(n+1)/2} (g A t^\lambda)^n  \\= \left(\frac{1}{\sqrt{i D_0 \tau }} - \frac{1}{a(t)} \right)^{-1},
\end{multline}
having performed the geometric series in the final step. This approximate form is equivalent to substituting $\mathbb{D}^{1/2, 1/\tau}\eta = e^{-t/\tau} D^{1/2}(e^{t/\tau} \eta) \approx \eta/\sqrt{\tau}$, which is a valid approximation when $t >\tau$ since then $e^{t/\tau}$ varies much faster in time than $\eta\lesssim\sqrt{t}$, such that the $D^{1/2}e^{t/\tau} \sim e^{t/\tau}/\sqrt{\tau}$ is the leading term.
Finally, we discuss the case $\lambda < 0$, where the power series given above would diverge at $t \rightarrow 0$. The solution is obtained in the same way as \eqref{eq:power-law-solution-app}, except that we write the differential form $-\eta/a = 1- D_t^{1/2}\eta /g $ as a Neumann series $ \eta = \sum_{n=0}^\infty b_n \tilde{K}^n[1]$, where $\tilde{K}[f]=  D_t^{1/2}[a f]/g$. Then the solution is given by
\begin{align}
    -\frac{\eta(t)}{a(t)} &=  \sum_{n= 0 }^\infty b_n (g A t^\lambda)^{-n} = \sum_{n=0}^\infty b_n \left(\frac{a(t)}{\sqrt{iD_0 t }} \right)^n, \\
    b_n(\lambda) &= \frac1{\Gamma(3/2) c_n(-\lambda)} = \prod_{k=1}^{n} \frac{\Gamma(1-k\lambda)}{\Gamma(\frac32 - k \lambda)}.
\end{align}
This form is regular at $t =0$ and allows us to read off the asymptotic behavior $\eta(t) =- a(t)$ for adiabatic drives of the scattering length $a(t) \ll \sqrt{iD_0 t}$, as discussed in the main text.

\bibliography{all}

\end{document}